\documentclass[preprint,12pt]{elsarticle}

\usepackage{amsmath,color}
\usepackage{longtable}
\usepackage{bbold}

\newcommand{\prefac}[1]{
    \ifcase #1
  \else
     1/(-\mu^2)^#1 \,    
  \fi
}

\newcommand{\G}{{\Gamma}}
\newcommand{\h}{$\displaystyle\phantom{\frac{1}{2}}$}

\newcommand{\pb}{{\bf p}}

\begin{document}

\begin{frontmatter}

  \title{                                                                                                              
    \vskip-3cm{\baselineskip14pt                                                                                             \centerline{\normalsize DESY 22-121\hfill ISSN 0418-9833}                                                                          \centerline{\normalsize July 2022\hfill}}                                                                      
    \vskip0.5cm
    Renormalization of three-quark operators 
    at two loops in the RI${}'$/SMOM scheme
  }    




      \author[mymainaddress]{Bernd A. Kniehl}
      \ead{kniehl@desy.de}
      \address[mymainaddress]{II.~Institut f\"ur Theoretische Physik, Universit\"at Hamburg,\\
        Luruper Chaussee 149, 22761 Hamburg, Germany}
      \author[mymainaddress]{Oleg L. Veretin}
      \ead{oleg.veretin@desy.de}
      \address[mymainaddress]{Institut f\"ur Theoretische Physik, Universit\"at Regensburg,\\
        Universit\"atsstra\ss{}e 31, 93040 Regensburg, Germany}

      \begin{abstract}
We consider the renormalization of the three-quark operators without
derivatives at next-to-next-to-leading order in QCD perturbation theory at the
symmetric subtraction point.
This allows us to obtain conversion factors between the $\overline{\rm MS}$
scheme and the regularization invariant symmetric MOM (RI/SMOM, RI${}'$/SMOM)
schemes.
The results are presented both analytically in $R_\xi$ gauge in terms of a set
of master integrals and numerically in Landau gauge.
They can be used to reduce the errors in determinations of baryonic
distribution amplitudes in lattice QCD simulations.
      \end{abstract}


      \end{frontmatter}

\section{Introduction}

Light cone distribution amplitudes (DAs) play an important r{\^o}le in the
analysis of hard exclusive reactions involving large momentum transfer from the
initial to the final state.
The cases of baryon asymptotic states have been considered already long ago
\cite{Lepage:1980fj,Efremov:1979qk,Chernyak:1983ej}.

The theoretical description of DAs is based on the relation of their moments
to matrix elements of local operators.\footnote{
  The three-quark operators relevant for the baryonic DAs have the general form
\begin{align}
  \epsilon^{ijk} \big( D_{\mu_1}\dots D_{\mu_l} \psi_1\big)^i_{\alpha_1}
  \big( D_{\mu_{l+1}}\dots D_{\mu_{l+m}} \psi_2\big)^j_{\alpha_2}
  \big( D_{\mu_{l+m+1}}\dots D_{\mu_{l+m+n}} \psi_3\big)^k_{\alpha_3} \,,
\end{align}  
where $i,j,k$ are color indices, $\mu_k$ are Lorentz indices, and $\alpha_l$
are spinor indices.}
Such matrix elements involve long-distance dynamics and, thus, cannot
be accessed via perturbation theory alone.

First estimates of the lower moments of the baryon DAs have been
obtained more than 30 years ago using QCD sum rules \cite{Chernyak:1984bm,King:1986wi,Chernyak:1987nu}.
An alternative way to access the moments is to calculate them from first
principles using lattice QCD.
Such studies for the nucleon DAs have a long
history \cite{Martinelli:1988xs,Gockeler:2008xv,QCDSF:2008qtn,Braun:2014wpa}.
More recently, this analysis has been extended to include the full
$\mbox{SU}(3)$ octet of baryons \cite{Bali:2015ykx}.

To renormalize the matrix elements on the lattice, 
the RI${}'$/SMOM scheme \cite{Sturm:2009kb} has been used in
Ref.~\cite{Bali:2015ykx}.
However, in order to embed lattice estimations of hadronic matrix
elements into the complex of other studies and to assure comparability, it
is necessary to present the result in the widely used $\overline{\mbox{MS}}$
scheme.
Since the RI${}'$/SMOM prescription can be used in both perturbative and
non-perturbative calculations, the conversion from the RI${}'$/SMOM to the
$\overline{\mbox{MS}}$ scheme can be evaluated perturbatively as a series in
the strong-coupling constant $\alpha_s(\mu)$ at some typical scale $\mu$ of the
order of a few GeV.
In our previous works \cite{Kniehl:2020nhw,Kniehl:2020sgo}, we have evaluated
the matching constants for the bilinear quark operators with up to two
derivatives and up to three loops.

In this paper, we perform the renormalization of the three-quark operators at
two loops for RI${}'$/SMOM kinematics, which allows for the conversion between
the RI${}'$/SMOM and $\overline{\mbox{MS}}$ schemes for the lowest moments of
the baryonic DAs.

As for baryonic operators, there are additional subtleties due to contributions
of evanescent operators that have to be taken into account.
In this work, we adopt the calculational scheme proposed in
Ref.~\cite{Krankl:2011gch}, which allows one to avoid the necessity of
additional finite renormalizations and the consideration of evanescent
operators.
Instead of contracting the operators with different Dirac
$\gamma$ matrices, we will consider the operators with open spinor indices.
The price that one has to pay is that one has a large number of different
spinor tensor structures.

This paper is organized as follows.
In Section~\ref{sec:two}, we introduce our notations and definitions.
In Section~\ref{sec:three}, we discuss the tensor decomposition and the
renormalizarion procedure.
In Section~\ref{sec:four}, we present a sample result, while our complete
result is provided in ancillary files submitted to the ArXiv along with this
paper.
In Section~\ref{sec:five}, we present our conclusions.
In \ref{sec:app}, we expose the relevant spin tensor structures. 

\section{Basic setup}
\label{sec:two}

\begin{figure}[h]
\centerline{\includegraphics[width=0.2\textwidth]{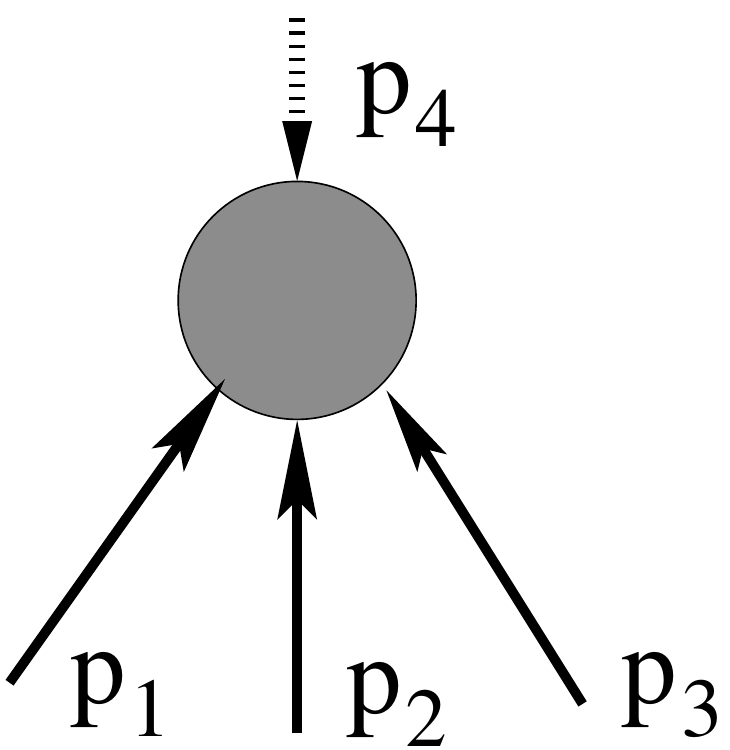}}
\caption{\label{fig:one}%
  Matrix element
  $\langle u^{b_1}_{\beta_1}(0) d^{b_2}_{\beta_2}(0) s^{b_3}_{\beta_3}(0)\bar{u}^{a_1}_{\alpha'_1}(p_1) \bar{d}^{a_2}_{\alpha'_2}(p_2) \bar{s}^{a_3}_{\alpha'_3}(p_3)
     \rangle$
     of a three-quark baryonic operator in momentum space.
     The momentum $p_4=-(p_1+p_2+p_3)$
     is the momentum coming into the operator.
}
\end{figure}

The basic object for the three-quark operators without derivatives located at
the origin is the amputated four-point function,
\begin{align}
\label{eq:def1}
H_{\beta_1\beta_2\beta_3,\alpha_1\alpha_2\alpha_3}(p_1,p_2,p_3) =& - \int d^4x_1 \, d^4x_2 \, d^4x_3
     e^{i(p_1x_1 + p_2x_2 + p_3x_3)}
     \epsilon^{b_1b_2b_3} \epsilon^{a_1a_2a_3} \nonumber\\
  & {}\times
     \langle u^{b_1}_{\beta_1}(0) d^{b_2}_{\beta_2}(0) s^{b_3}_{\beta_3}(0)
     \bar{u}^{a_1}_{\alpha'_1}(p_1) \bar{d}^{a_2}_{\alpha'_2}(p_2) \bar{s}^{a_3}_{\alpha'_3}(p_3)
     \rangle \nonumber\\
  & {}\times
     G_2^{-1}(p_1)_{\alpha'_1\alpha_1}
     G_2^{-1}(p_2)_{\alpha'_2\alpha_2}
     G_2^{-1}(p_3)_{\alpha'_3\alpha_3} \,,
\end{align}  
where all quantities are to be understood as Euclidean.
The quark flavors are called $u$, $d$, and $s$, but the only essential feature
is that they are all different.
All masses are supposed to vanish.
$\alpha_i$ and $\beta_j$ are spinor indices, $a_k$ and $b_l$ are color indices
in the fundamental representation, and $p_m$ are the external momenta.
The matrix element of the three-quark operator is shown schematically in
Fig.~\ref{fig:one}.

The two-point function $G_2(p)$ required for the amputation of the external
legs is defined by
\begin{align}
  \delta^{a'a}\,G_2(p)_{\alpha'\alpha} = \int d^4x \, e^{i px}
  \langle u^{a'}_{\alpha'}(0) \bar{u}^a_\alpha(x)  \rangle \,.
\end{align}  

To compute the conversion factor for a particular multiplet of operators, the
amputated four-point function (\ref{eq:def1}) has to be contracted with
coefficients
\begin{align}
C^{\beta_1\beta_2\beta_3,\alpha_1\alpha_2\alpha_3} \,,
\end{align}
which can be easily provided.
Notice, however, that these coefficients refer to a particular representation
of the Dirac matrices.


In the following, we use the kinematics that was adopted in the analysis of
Ref.~\cite{Bali:2015ykx},
\begin{align}
  p_1^2 &= p_2^2 = p_3^2 = (p_1+p_2+p_3)^2 = \mu^2 \,,\\
  p_1\cdot p_2 &= p_3\cdot p_1 = -\frac{1}{2}\mu^2 \,,\\
  p_2\cdot p_3 &= 0 \,,
\end{align}
where $\mu$ is some euclidean point that fixes the SMOM subtraction point.


\section{Tensor decomposition and projection}
\label{sec:three}

As was already mentioned in the Introduction, we renormalize Eq.~(\ref{eq:def1})
without contracting the spinor indices and projecting on some particular
baryonic currents.
For this purpose, let us decompose the tensor in Eq.~(\ref{eq:def1}) as
\begin{align}
\label{eq:decomposition}  
  H_{\beta_1\beta_2\beta_3,\alpha_1\alpha_2\alpha_3} (p_1,p_2,p_3) &=
  \sum\limits_{n=1}^N T_{n,\beta_1\beta_2\beta_3,\alpha_1\alpha_2\alpha_3} (p_1,p_2,p_3)
  \,f_n\Big(\{p_ip_j\}\Big) \,,
\end{align}
where $T_n$ are spin tensor structures and $f_n$ are scalar form factors.
The explicit construction of these structures is discussed in \ref{sec:app}.
The form factors $f_n$ generally depend on six kinematic invariants,
$p_1^2$, $p_2^2$, $p_3^2$, $p_1\cdot p_2$, $p_2\cdot p_3$, and $p_3\cdot p_1$.
In the following discussion, we omit spinor indices and arguments and simply
write
\begin{align}
\label{eq:decomposition1}  
  H = \sum\limits_{n=1}^N T_n\,f_n \,.
\end{align}
The upper limit $N$ of summation in Eqs.~(\ref{eq:decomposition}) and
(\ref{eq:decomposition1}) is the number of the linearly independent spin tensor
structures.
It depends on the number of loops.
We also have to distinguish between the decompositions in $d$ and four
dimensions.
In $d$ dimensions, the number of independent structures is larger, owing to the
presence of evanescent operators.
The values $N$ of independent form factors through two loops are given in
Table~\ref{tab:N}.
\begin{table}[h]
\centerline{
\begin{tabular}{|c|c|c|c|}
 \hline
  \# of loops & ~~~~~0~~~~~ & ~~~~~1~~~~~ & ~~~~~2~~~~~ \\
 \hline
  $N$ (in  $d$ dimensions) & 1 & 67 & 581 \\
 \hline
  $N$ (in 4 dimensions) & 1 & 64 & 247 \\  
 \hline
\end{tabular}
}
\caption{Number of form factors for different numbers of loops in $d$ and four
  dimensions.}
  \label{tab:N}
\end{table}

Let us introduce the following notation.
If $X_{\beta_1\beta_2\beta_3,\alpha_1\alpha_2\alpha_3}$ is an object with six spinor
indices, we denote by $\mbox{tr}_3(X)$ the trace over three pairs of indices,
i.e.,
\begin{align}
\label{eq:tr}  
\mbox{tr}_3 (X) = \sum_{\alpha_1,\alpha_2,\alpha_3=1}^4
    X_{\alpha_1\alpha_2\alpha_3,\alpha_1\alpha_2\alpha_3} \,.
\end{align}
Using this definition, we can introduce the $N\times N$ symmetric matrix
\begin{align}
\label{eq:M}  
  M_{kn} = \mbox{tr}_3 (T_k T_n) \,,
\end{align}
where $T_j$ are the spin tensor structures from Eqs.~(\ref{eq:decomposition})
and (\ref{eq:decomposition1}).
Then, the projectors on the form factors $f_j$ take the form
\begin{align}
\label{eq:P}  
  P_{l} = \sum_{k=1}^N M^{-1}_{lk} \, T_k \,,
\end{align}
where $M^{-1}$ is the inverse matrix, and we obviously have
\begin{align}
\label{eq:proj}  
  f_l = \mbox{tr}_3\left(P_{l}H\right) \,.
\end{align}
The use of Eqs.~(\ref{eq:P}) and (\ref{eq:proj}) for unrenormalized amplitudes
is delicate within dimensional regularization, since the projectors $P_l$
depend nontrivially on the dimension $d$.
A better way is to first renormalize the amplitude $H$ and then use the
projectors in four dimensions.
In order to achieve this, we construct $N$ scalar amplitudes $A_k$ as
\begin{align}
\label{eq:A}  
  A_k = \mbox{tr}_3 (T_k H) \,,  \qquad k=1,\dots,N \,.
\end{align}
After renormalization of all $A_k$ amplitudes in the $\overline{\mbox{MS}}$
scheme, the form factors
can be obtained as\footnote{
  Indeed, we have
  $\sum_k M^{-1}_{lk} A_k = \sum_{k,n} M^{-1}_{lk} M_{kn} f_n =\sum_{n} \delta_{ln} f_n= f_l$.
  }
\begin{align}
\label{eq:f}  
  f_l = \sum\limits_{k=1}^N M^{-1}_{lk} \, A_k \,,
\end{align}
where $M^{-1}$ is now taken in four dimensions.
In this limit, all elements of $M^{-1}$ are just rational numbers.


However, in four dimensions, we cannot apply the formula in Eq.~(\ref{eq:f})
directly, since the determinant of the matrix $M_{lk}$ is then zero.
This may be understood from Table~\ref{tab:N} by observing that the number of
independent structures in four dimensions is less than in $d$ dimensions.
In this case, we need to solve the system (in matrix notation)
\begin{align}
  M\vec{A}=\vec{f} \,,
  \label{eq:sys}
\end{align}
where $\vec{f}=(f_1,\dots,f_N)^T$ etc.

The system~(\ref{eq:sys}) is over-determined, but consistent by construction.
We find the solution for $\vec{f}$ in the form
\begin{align}
  \vec{f} = \vec{f}_0 + \sum_{j=0}^{N_d-N_4} C_j \vec{y}_j \,, 
\label{eq:solution}
\end{align}
where $\vec{f}_0$ is some particular solution of the system~(\ref{eq:sys}), the
vectors $\vec{y}_j$ form a basis of the $N_d-N_4=334$ dimensional null space of
the matrix $M$, and $C_j$ are arbitrary constants.

After renormalization, we have 581 two-loop form factors $f_n$ in four
dimensions, 247 of which are linearly independent.
We have calculated all of them analytically in $R_\xi$ gauge in terms of a set
of complicated master integrals, which we have evaluated numerically.


\section{Results}
\label{sec:four}

Because of their large number, we refrain from listing the renormalized
two-loop form factors $f_n$ here, but supply them in ancillary files submitted
to the ArXiv along with this manuscript.
Specifically, we present our analytic results in $R_\xi$ gauge in the form of
Eq.~(\ref{eq:solution}) including explicit expressions for the constants $C_i$,
and our numerical results in Landau gauge ($\xi=0$) for $C_i=0$.

To illustrate the structure and typical size of the corrections, we present
here, in numerical form, the two-loop form factor $f_1$, corresponding to the
structure $\Gamma_0\otimes\Gamma_0\otimes\Gamma_0$, in $R_\xi$ gauge:
\begin{align}
  \frac{f_1}{f_{1,\mbox{Born}}} ={}&1 + a (0.6204053 + 0.595702\,\xi) \nonumber\\
  &{}+ a^2 [ 10.45  + 3.59\,\xi + 1.42\,\xi^2 \pm 0.03
   - (0.689 \pm 0.001) n_f] \,,
\end{align}
where $f_{1,\mbox{Born}}=\epsilon^{ijk}\epsilon^{ijk}=6$ is the Born result,
$a=\alpha_s/\pi$, $n_f$ is the number of light quark flavors, and $\xi$ is the
gauge parameter.

\section{Conclusion}
\label{sec:five}

In this work, we have established a framework for the evaluation of the
corrections to the baryonic current without derivatives through the two-loop
order.
The main difficulty in the study of the baryonic operators is the presence of
evanescent operators that mix under renormalization with the physical operators.
This leads to a large mixing matrix and the necessity for finite
renormalizations.
On the other hand, if we use the open-indices approach, there is no ambiguity
in the interpretation in the $\overline{\mbox{MS}}$ scheme.
Exploiting this observation, we have evaluated all the form factors appearing
through two loops and presented them in a numerical form that is ready for use
in lattice QCD simulations.\\

{\it Acknowledgments:}
We are grateful to Vladimir M.~Braun and Meinulf~G\"ockeler, and
Alexander N.~Manashov for fruitful discussions.
O.L.V. is grateful to the University of Hamburg for the warm hospitality.
This work was supported in part by the German Research Foundation DFG through
Research Unit FOR~2926 ``Next Generation Perturbative QCD for Hadron Structure:
Preparing for the Electron-Ion Collider" with Grant No.~KN~365/13-1.


\appendix
\section{Spin tensor structures}
\label{sec:app}

In this Section, we explicitly enumerate all linearly independent spin tensor
structures $T_n$ through two loops in $d$ dimensions.
All tensors $T_n$ are represented as a tensor products of three Dirac
structures, as
\begin{align}
T_{\alpha_1\alpha_2\alpha_3,\beta_1\beta_2\beta_3} =
    \G_{\alpha_1\beta_1}\otimes\G_{\alpha_2\beta_2}\otimes\G_{\alpha_3\beta_3} \,.
\end{align}
The building blocks $\G$ are anti-symmetric products of Dirac $\gamma$ matrices,
\begin{align}
  \G_{0} &= \mathbb{1} \,, \\
  \G_{\mu_1\mu_2} &= \frac{1}{2!} \gamma_{[\mu_1}\gamma_{\mu_2]}  \,, \\
  \G_{\mu_1\mu_2\mu_3\mu_4} &= \frac{1}{4!} \gamma_{[\mu_1}\gamma_{\mu_2}
         \gamma_{\mu_3}\gamma_{\mu_4]}  \,, 
\end{align}
where $\mathbb{1}$ is the unit Dirac matrix and square brackets $[\dots]$
denote antisymmetrisation.
Notice that Dirac structures with odd numbers of Dirac matrices do not
appear in our calculation. 

We also introduce the following notation for the contraction of a vector and a
tensor (Schoonship notation)
\begin{equation}
  p^\mu \G_{\dots\mu\dots}=\G_{\dots p\dots}\,.
\end{equation}  
Furthermore, we introduce the following wild-cards:
$\pb$ can take one of $p_1,p_2,p_3$, $\pb\pb$ can take one of
$p_1p_2,p_2p_3,p_3p_1$, and $\pb\pb\pb$ stands for $p_1p_2p_3$.

For the sake of systematics, we assign to each tensor structure a signature,
which is an ordered triplet of the numbers 0, 2, and 4 of $\gamma$ matrices
appearing in each $\Gamma$ factor, and a number $[p]$ counting the overall
appearances of momenta.
Furthermore, we distinguish between symmetric and non-symmetric structures.
The symmetric structures do not have co-partners arising under the change of
order of the $\Gamma$ factors in the tensor products, while the non-symmetric
ones do.
So, the numbers of non-symmetric structures should be multiplied by 3. 
In Tables~\ref{tab:sym} and \ref{tab:nonsym}, we systematically list the
symmetric and non-symmetric tensor structures, respectively, and specify the
number (\#) of entities for each signature and each value of $[p]$.
We also give the total number (\#\#) of entities for each signature.

\begin{table}
\centerline{
\begin{tabular}[h!]{|c|c|c|c|c|}
 \hline
  signature &  $[p]$ & tensor structure & \# & \#\# \\
 \hline
 000 & 0 & \h $\G_{0}\otimes\G_{0}\otimes\G_{0}$ & 1  & 1 \\
   \hline
 222 & 0 & \h $\G_{\mu_1\mu_2}\otimes\G_{\mu_2\mu_3}\otimes\G_{\mu_3\mu_1}$ & 1 & \\
 222 & 6 & \h $\prefac{3}\G_{\pb\pb}\otimes\G_{\pb\pb}\otimes\G_{\pb\pb}$ & 27 & 28 \\
 \hline
\end{tabular}
}
\caption{Symmetric structures ordered according to their signatures and values
  of $[p]$, numbers \# of entities for given signature and value of $[p]$, and
  total numbers \#\# of entities for given signature.}
\label{tab:sym}
\end{table}

\newpage



\begin{center}
\begin{longtable}[h!]{|c|c|c|c|c|}
 \hline
  signature & $[p]$ & tensor structure & \# & \#\# \\
 \hline
 200 & 2 & \h $\prefac{1}\G_{\pb\pb}\otimes\G_{0}\otimes\G_{0}$ & 3 & 3 \\
   \hline
 220 & 0 & \h $\prefac{0}\G_{\mu_1\mu_2}\otimes\G_{\mu_1\mu_2}\otimes\G_{0}$ & 1 & \\
 220 & 2 & \h $\prefac{1}\G_{\pb\mu_1}\otimes\G_{\pb\mu_1}\otimes\G_{0}$     & 9 & \\
 220 & 4 & \h $\prefac{2}\G_{\pb\pb}\otimes\G_{\pb\pb}\otimes\G_{0}$         & 9 & 19 \\
   \hline
 222 & 2 & \h $\prefac{1}\G_{\pb\mu_1}\otimes\G_{\pb\mu_2}\otimes\G_{\mu_1\mu_2}$ & 9  & \\
 222 & 2 & \h $\prefac{1}\G_{\pb\pb}\otimes\G_{\mu_1\mu_2}\otimes\G_{\mu_1\mu_2}$ & 3  & \\
 222 & 4 & \h $\prefac{2}\G_{\pb\pb}\otimes\G_{\pb\mu_1}\otimes\G_{\pb\mu_1}$     & 27 & 39 \\
   \hline
 402 & 2 & \h $\prefac{1}\G_{\pb\pb\mu_1\mu_2}\otimes\G_{0}\otimes\G_{\mu_1\mu_2}$  & 3 & \\
 402 & 4 & \h $\prefac{2}\G_{\pb\pb\pb\mu_1}\otimes\G_{0}\otimes\G_{\pb\mu_1}$   & 3 & 6 \\
   \hline
 420 & 2 & \h $\prefac{1}\G_{\pb\pb\mu_1\mu_2}\otimes\G_{\mu_1\mu_2}\otimes\G_{0}$  & 3 & \\
 420 & 4 & \h $\prefac{2}\G_{\pb\pb\pb\mu_1}\otimes\G_{\pb\mu_1}\otimes\G_{0}$   & 3 & 6 \\
   \hline
 440 & 0 & \h $\prefac{0}\G_{\mu_1\mu_2\mu_3\mu_4}\otimes\G_{\mu_1\mu_2\mu_3\mu_4}\otimes\G_{0}$ & 1 & \\
 440 & 2 & \h $\prefac{1}\G_{\pb\mu_1\mu_2\mu_3}\otimes\G_{\pb\mu_1\mu_2\mu_3}\otimes\G_{0}$     & 9 & \\
 440 & 4 & \h $\prefac{2}\G_{\pb\pb\mu_1\mu_2}\otimes\G_{\pb\pb\mu_1\mu_2}\otimes\G_{0}$         & 9 & \\
 440 & 6 & \h $\prefac{3}\G_{\pb\pb\pb\mu_1}\otimes\G_{\pb\pb\pb\mu_1}\otimes\G_{0}$             & 1 & 20 \\
   \hline
 422 & 0 & \h $\prefac{0}\G_{\mu_1\mu_2\mu_3\mu_4}\otimes\G_{\mu_1\mu_2}\otimes\G_{\mu_3\mu_4}$ & 1  & \\
 422 & 2 & \h $\prefac{1}\G_{\pb\mu_1\mu_2\mu_3}\otimes\G_{\pb\mu_1}\otimes\G_{\mu_2\mu_3}$     & 9  & \\
 422 & 2 & \h $\prefac{1}\G_{\pb\mu_1\mu_2\mu_3}\otimes\G_{\mu_2\mu_3}\otimes\G_{\pb\mu_1}$     & 9  & \\
 422 & 2 & \h $\prefac{1}\G_{\pb\pb\mu_1\mu_2}\otimes\G_{\mu_2\mu_3}\otimes\G_{\mu_3\mu_1}$     & 3  & \\
 422 & 4 & \h $\prefac{2}\G_{\pb\pb\mu_1\mu_2}\otimes\G_{\pb\mu_1}\otimes\G_{\pb\mu_2}$         & 27 & \\
 422 & 4 & \h $\prefac{2}\G_{\pb\pb\mu_1\mu_2}\otimes\G_{\pb\pb}\otimes\G_{\mu_1\mu_2}$         & 9  & \\
 422 & 4 & \h $\prefac{2}\G_{\pb\pb\mu_1\mu_2}\otimes\G_{\mu_1\mu_2}\otimes\G_{\pb\pb}$         & 9  & \\
 422 & 4 & \h $\prefac{2}\G_{\pb\pb\pb\mu_1}\otimes\G_{\pb\mu_2}\otimes\G_{\mu_1\mu_2}$         & 3  & \\
 422 & 4 & \h $\prefac{2}\G_{\pb\pb\pb\mu_1}\otimes\G_{\mu_1\mu_2}\otimes\G_{\pb\mu_2}$         & 3  & \\
 422 & 6 & \h $\prefac{3}\G_{\pb\pb\pb\mu_1}\otimes\G_{\pb\pb}\otimes\G_{\pb\mu_1}$           & 9  &   \\
 422 & 6 & \h $\prefac{3}\G_{\pb\pb\pb\mu_1}\otimes\G_{\pb\mu_1}\otimes\G_{\pb\pb}$           & 9  & 91\\
 \hline
 \caption{Non-symmetric structures. The meaning of the columns is the same as
   in Table \ref{tab:sym}.}
\label{tab:nonsym}
\end{longtable}
\end{center}


\bibliographystyle{plain}
\bibliography{references}

\end{document}